\documentstyle[prl,aps,preprint,axodraw]{revtex}

\def\btabl{\begin{table}}   \def\etabl{\end{table}}
\def\bea{\begin{eqnarray}}   \def\eea{\end{eqnarray}}
\def\bnn{\begin{eqnarray*}}   \def\enn{\end{eqnarray*}}
\def\beq{\begin{equation}}   \def\eeq{\end{equation}}  
\def\btabu{\begin{tabular}}   \def\etabu{\end{tabular}}
\def\bec{\begin{displaymath}} \def\eec{\end{displaymath}}
\def\nn{\nonumber}
\def\eqref#1{(\ref{#1})}

\renewcommand{\baselinestretch}{1.2}
\begin{document}
\newcommand{\bfig}{\begin{center}\begin{picture}}
\newcommand{\efig}[1]{\end{picture}\\{\small #1}\end{center}}
\newcommand{\flin}[2]{\ArrowLine(#1)(#2)}
\newcommand{\wlin}[2]{\DashLine(#1)(#2){3}}
\newcommand{\zlin}[2]{\DashLine(#1)(#2){5}}
\newcommand{\glin}[3]{\Photon(#1)(#2){2}{#3}}
\newcommand{\lin}[2]{\Line(#1)(#2)}
\newcommand{\sof}{\SetOffset}
\draft

\preprint{\vbox{\baselineskip=13pt
\rightline{CERN-TH/98-182}
\rightline{LPTHE Orsay-98/29}
\rightline{hep-ph/9806383}}}
\title{Inelastic photon--neutrino interactions \\
using an effective Lagrangian}
\author{A. Abada, 
J. Matias  and R. Pittau \footnote{
e-mail: abada@mail.cern.ch, matias@mail.cern.ch, pittau@mail.cern.ch.}} 
\vskip -0.5cm
\vspace{-0.5cm}
\address{{\small 
Theory Division, CERN, CH-1211 Geneva 23, Switzerland.}}

\vskip -1cm

\maketitle \begin{abstract}  
We justify the feasibility of substituting a photon leg by a neutrino current
 in the Euler--Heisenberg
 Lagrangian to obtain an effective Lagrangian for the 
process  $\gamma\nu\to\gamma\gamma \nu$ and its crossed
 reactions. 
  We establish the link between these processes and the four-photon scattering 
  in both the Standard Model and the effective theory.
As an application, we  compute the processes $\gamma\nu\to\gamma\gamma\nu$ and  
$\gamma\gamma\to\gamma{\nu}\bar\nu$, 
give their polarized cross sections, and show how to use the $\gamma\gamma\to\gamma\gamma$ 
results as a check. 
We settle the question about the disagreement between two computations
 in the literature concerning the reaction
$\gamma\gamma\to \gamma \nu\overline\nu$.\\ \\
Pacs numbers: 13.10.+q, 13.15.+g, 14.70.Bh, 13.88.+e, 95.30.Cq.
\end{abstract}

\vspace{3.4cm}

\leftline{} 
\leftline{CERN-TH/98-182}
\leftline{June 1998}  

\renewcommand{\baselinestretch}{1.2}

\newpage \section{Introduction}
Processes involving neutrinos play an important role in astrophysics.
For example, during the evolution of a star, neutrino emission 
carries energy away from the entire volume of the star, 
while, because of short range interactions, the compact (dense) central
part is opaque for photons, which remain trapped inside. In that sense, 
contrarily to the neutrino, the photon is an ``amnesic'' particle. 

In particular, despite their small cross section, 
low-energy photon--neutrino processes are potentially 
of interest in stellar evolution as well as in cosmology.
Typical examples are the processes
\begin{eqnarray}
\gamma\nu   &\to& \gamma\nu       \label{eq1}     \\
\gamma\gamma&\to& \nu\overline\nu \label{eq2}\\
\nu\overline\nu&\to&\gamma\gamma  \label{eq3}\,.
\end{eqnarray}
Nevertheless, the cross section for such reactions is too small. In fact,
due to the Yang theorem \cite{Yang}, which forbids two photons to couple
in a $J=1$ state, their amplitude is exactly zero 
to order $G_F$ \cite{Gell} in the Standard Model (SM) and
they are suppressed by powers of 
$\omega/M_W$, where $\omega$ is the centre-of-mass energy of the photon
and $M_W$ the $W$ boson mass. In case of massive neutrinos, a
suppression factor $m_\nu/M_W$ is also present, where $m_\nu$ 
is the neutrino mass.
For massless neutrinos, the first non-zero contribution 
is always of order $1/M_W^4$, and the SM cross 
sections have been shown to be negligibly small in ref. \cite{dic2}.
 
On the contrary, five-leg processes involving two neutrinos 
and three photons, such as
\begin{eqnarray}
\gamma\nu    &\to& \gamma\gamma\nu     \label{eq4}\\
\gamma\gamma &\to& \gamma\nu\bar\nu    \label{eq5}\\
\nu\bar\nu   &\to& \gamma\gamma\gamma  \label{eq6}\,, 
\end{eqnarray}
are not constrained by Yang's theorem. 
Moreover, the extra $\alpha$ in the cross section is compensated 
by an interchange of the $\omega/M_W$ suppression by an 
$\omega/m_e$ enhancement.  

  Recently, Dicus and Repko derived an effective Lagrangian for five-leg
photon--neutrino interactions \cite{dic3}. They based their derivation on the
Euler--Heisenberg Lagrangian \cite{EH}. After Fierz rearranging, taking 
the large mass gauge boson limit, and applying Furry's theorem, they found that
 the amplitude
of the five-leg photon--neutrino process reduces to a four-photon amplitude
with one photon field replaced by the neutrino current. 
 Using such a Lagrangian, they calculated processes 
(\ref{eq4}), (\ref{eq5}) and (\ref{eq6})
for energies below the threshold for $e^+ e^-$ pair production, 
showing that the energy dependence is $ \omega^{10}$. 
By extrapolating this result
beyond the range of validity of the effective Lagrangian,
the resulting cross sections are of the order of $10^{-52}\,{\rm cm^2}$,
for $\omega\sim 1$ MeV. 
This is to be compared with process (\ref{eq1}),
whose cross section is of the order of $10^{-65}{\rm cm^2}$, for 
$\omega\sim 1$ MeV \cite{dic2}. 
 
Basing their work on ref. \cite{dic3}, Harris, Wang and Teplitz  
\cite{teplitz} investigated to which extent five-leg neutrino processes affect 
the supernova dynamics, in case the results of ref.\cite{dic3} could 
indeed be extrapolated above $1$ MeV.
They estimated that, in order to fit with the data of SN87A, 
the cross sections of  reaction (\ref{eq4}) 
should have the behaviour 
$10^{-52}({\omega/1\ {\mathrm{MeV}}})^{\sim 8.4}\,{\rm cm^2}$ 
for $\omega$ of the order of few MeV.
 
 Furthermore, the computation of process (\ref{eq5}) in ref. \cite{dic3}
disagrees with the result obtained in 1963 
by Hieu and Shabalin \cite{shab}. 

 Given the above scenario, the aim of this work is threefold.
First, we would like to demonstrate, step by step, in a somehow
pedagogical
way, the derivation of 
the effective theory for inelastic photon--neutrino processes  
from the effective Euler--Heisenberg theory describing
elastic photon--photon interactions. Secondly, we want to settle the 
question of the disagreement between the 
calculations of refs. \cite{dic3} and \cite{shab} 
for process (\ref{eq5}), which is an important 
energy-loss mechanism in the stellar evolution. 
The conclusion is that our result agrees with 
that of ref. \cite{dic3}. Thirdly, as a cross-check of ref. \cite{dic3},  
we compute the direct process $\gamma\nu\to\gamma\gamma\nu$,  
explicitly giving all polarized cross sections in the effective
theory. 
All calculations in this paper are performed with massless neutrinos.

 It goes without saying that computing processes 
(\ref{eq4}), (\ref{eq5}) and (\ref{eq6}) exactly in the SM, 
for energies above 
the $e^+e^-$ production threshold, is of extreme interest
so as to precisely settle their role in astrophysics.  
Such a computation is under way \cite{nous}, and, when completed,
will give cross sections valid for all energies up to $\omega< M_W$,
therefore setting the real range of validity of the 
effective theory.

The outline of the paper is as follows.
In section II, we derive the effective five-leg Lagrangian from the SM. 
In section III, we compute the polarized and the total cross
section for $\gamma\gamma\to \gamma {\nu}\bar\nu$.  
Finally, in section IV, we deal with the process 
$\gamma{\nu}\to \gamma \gamma\nu$, whose 
polarized cross sections are explicitly presented in the appendix. 
 
\section{Derivation of the effective Lagrangian}\label{Lagrangian}

Our starting point is the leading term of the Euler--Heisenberg 
Lagrangian \cite{EH}, describing the photon--photon interaction of 
fig. 1a:
\bea 
{\cal L}_{{\mathrm E-H}}={\alpha^2\over 180 m_e^4}
\left [ 5\left(F_{\mu\nu}F^{\mu\nu}\right)^2 
-14 F_{\mu\nu}F^{\nu\lambda}F_{\lambda\rho}F^{\rho\mu}\right ] + {\cal
O}(\alpha^3).
\label{e-h}
\eea
In this equation, $\alpha$ is the QED fine structure constant,
$m_e$ the electron mass and $F_{\mu\nu}$ the  photon field-strength
tensor. 
${\cal L}_{{\mathrm E-H}}$ can be obtained by matching,
in the large $m_{e}$ limit, the exact result  
with all possible operators compatible with the symmetries,
at a given order in the momenta and powers of $1/M_{W}^{2}$.
\noindent

\bfig(300,120)
\sof(-50,20)
\GOval(110,40)(4,4)(0){0}
\glin{110,0}{110,40}{5}
\glin{110,40}{65,40}{5}
\glin{110,40}{110,85}{5}
\glin{110,40}{155,40}{5}  
\Text(115,12)[l]{$\gamma$}   
\Text(115,70)[l]{$\gamma$}
\Text(77,36)[t]{$\gamma$} 
\Text(143,36)[t]{$\gamma$}
\Text(60,40)[r]{$p_1,~\epsilon_1$} 
\Text(160,40)[l]{$p_3,~\epsilon_3$}
\Text(110,90)[b]{$p_2,~\epsilon_2$}
\Text(110,-2)[t]{$p_4,~\epsilon_4$}
\Text(150,87)[bl]{$(a)$}
\sof(140,20)
\glin{110,0}{110,20}{3}
\flin{110,20}{90,40}
\flin{90,40}{110,60}
\flin{110,60}{130,40}
\flin{130,40}{110,20}
\glin{90,40}{65,40}{3}
\glin{110,60}{110,85}{3}
\glin{130,40}{155,40}{3}  
\Text(115,12)[l]{$\gamma$}   
\Text(98,55)[r]{$e$}
\Text(115,70)[l]{$\gamma$}
\Text(77,36)[t]{$\gamma$} 
\Text(143,36)[t]{$\gamma$}
\Text(60,40)[r]{$p_1,~\epsilon_1$} 
\Text(160,40)[l]{$p_3,~\epsilon_3$}
\Text(110,90)[b]{$p_2,~\epsilon_2$}
\Text(110,-2)[t]{$p_4,~\epsilon_4$}
\Text(150,87)[bl]{$(b)$}
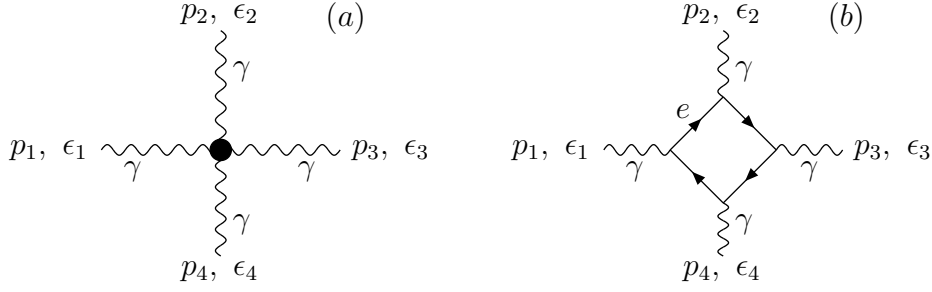
\efig{Fig. 1:  Four-photon interaction in the effective Lagrangian 
(a)  and in the SM (b).}

In order to find the relation between the effective Lagrangian of eq.
 (\ref{e-h}) and the one describing processes
(\ref{eq4}), (\ref{eq5}) and (\ref{eq6}), we write down the corresponding
amplitudes in the SM,  and show
under which approximations they are equivalent, up to a global factor.   

At the one-loop and at  leading order in $1/M_W^2$,
reactions (\ref{eq4}), (\ref{eq5}) and (\ref{eq6})
are given by the diagrams of fig. 2, plus all possible permutations 
of the photon legs.
The scale is determined by the mass of the fermion running in the
loop, so that, at low energies, the leading contribution
is given by the mass of the electron. Amplitudes involving 
higher mass particles, such as $\mu$, $\tau$,
gauge bosons $G$, or hadrons $H$ 
(at these energies quarks combine into hadrons),
are suppressed by powers of $1/M_i$, with $i=\mu,\ \tau, G,H$. 
It is precisely the appearance of $m_{e}$ as the scale,
instead of $M_{W}$ (which turns out to be the scale that controls  reactions
(\ref{eq1}), (\ref{eq2}) and (\ref{eq3})),
that makes these processes relevant at energies of the
order of few MeV. 

\bfig(300,120)
\sof(-50,20)
\flin{40,0}{110,0}  \flin{110,0}{180,0}
\wlin{110,0}{110,20}\
\flin{110,20}{90,40}
\flin{90,40}{110,60}
\flin{110,60}{130,40}
\flin{130,40}{110,20}
\glin{90,40}{65,40}{3}
\glin{110,60}{110,85}{3}
\glin{130,40}{155,40}{3}
\Text(63,-5)[t]{$p_4~~~~\nu$}
\Text(155,-5)[t]{$\nu~~~~p_5$}
\Text(115,12)[l]{$Z$}
\Text(98,55)[r]{$e$}
\Text(115,70)[l]{$\gamma$}
\Text(77,36)[t]{$\gamma$}
\Text(143,36)[t]{$\gamma$}
\Text(60,40)[r]{$p_1,~\epsilon_1$}
\Text(160,40)[l]{$p_3,~\epsilon_3$}
\Text(110,90)[b]{$p_2,~\epsilon_2$}
\Text(150,87)[bl]{$(a)$}
\sof(140,20)
\flin{40,0}{80,0}  \wlin{80,0}{140,0} \flin{140,0}{180,0}
\flin{80,0}{90,20} \flin{90,20}{110,32}
\flin{110,32}{130,20} \flin{130,20}{140,0}
\glin{90,20}{67,37}{3}
\glin{110,32}{110,62}{3}
\glin{130,20}{153,37}{3}
\Text(63,-5)[t]{$p_4~~~~\nu$}
\Text(155,-5)[t]{$\nu~~~~p_5$}
\Text(110,4)[b]{$W$}
\Text(98,33)[r]{$e$} 
\Text(115,50)[l]{$\gamma$}
\Text(75,23)[t]{$\gamma$} 
\Text(145,23)[t]{$\gamma$}
\Text(65,45)[r]{$p_1,~\epsilon_1$}
\Text(155,45)[l]{$p_3,~\epsilon_3$}
\Text(110,70)[b]{$p_2,~\epsilon_2$}
\Text(150,87)[bl]{$(b)$}
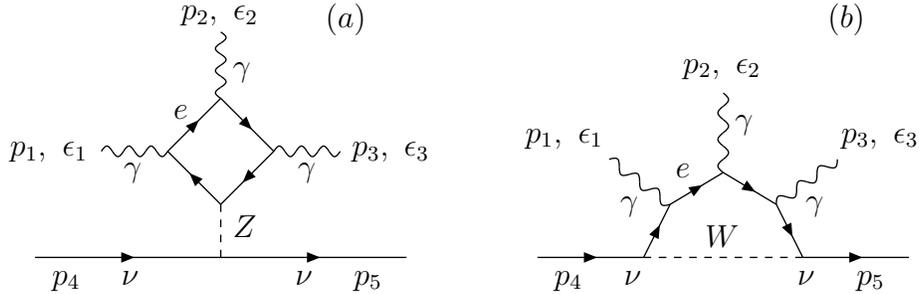
\efig{Fig. 2: SM leading diagrams contributing to five-leg 
photon--neutrino processes.}

We denote by $A_{ijk}$ and $B_{ijk}$ the contributions coming 
from diagrams (a) and (b) in fig.2, respectively. With this notation,
$i,j,k$ label a particular permutation of the photon legs. 
Therefore, the total amplitude reads
\bea
{\cal A}_P^{\rm SM} &=&\epsilon^{\alpha}(\vec{P}_{1},\lambda_{1})
\epsilon^{\beta}(\vec{P}_{2},\lambda_{2})
\epsilon^{\gamma}(\vec{P}_{3},\lambda_{3}) [
(A_{123}^{\alpha
\beta
\gamma}+A_{321}^{\alpha \beta
\gamma})+(A_{132}^{\alpha \beta
\gamma}+A_{231}^{\alpha \beta \gamma})+(A_{213}^{\alpha \beta
\gamma}+A_{312}^{\alpha \beta \gamma})+
\nn \\ &&(B_{123}^{\alpha
\beta \gamma}+B_{321}^{\alpha \beta
\gamma})+(B_{132}^{\alpha \beta \gamma}+B_{231}^{\alpha 
\beta \gamma})+(B_{213}^{\alpha \beta \gamma}+B_{312}^{\alpha \beta
\gamma})]\,, \eea
where $\epsilon^{\alpha}(\vec{P}_{i},\lambda_{i})$ are 
the polarization vectors associated with the $i-{\rm th}$ photon.
We grouped the amplitudes because, as we shall see, 
it is convenient to consider the diagrams in pairs. 
For instance, the amplitudes for the first two diagrams are
\bea
A_{123}^{\alpha \beta \gamma} &=& -{(g s_{W})}^3 {\left( {g \over 2
c_{W}} \right)}^2 \!
\sum_{\tau=+,-} v_{e}^{\tau}\Gamma_{\mu} \left({1
\over \Delta_{Z}}
 \right) {1 \over {(2 \pi)}^{4}} \int dq^{n} 
{\rm \,Tr\,}\left[\gamma^{\mu} w^{\tau} {1
\over
Q_{23}^{-}} \gamma^{\gamma} {1 \over Q_{2}^{-}} \gamma^{\beta} {1 \over
Q_{0}^{-}}
\gamma^{\alpha} {1 \over Q_{1}^{-}} \right] \nonumber \\ 
A_{321}^{\alpha \beta \gamma} &=& -{(g s_{W})}^3 {\left( {g \over 2
c_{W}} \right)}^2\!
\sum_{\tau=+,-} v_{e}^{\tau}\Gamma_{\mu} \left({1
\over \Delta_{Z}}
 \right) {1 \over {(2 \pi)}^{4}} \int dq^{n} 
{\rm \,Tr\,}
\left[\gamma^{\mu} w^{\tau} {1   
\over
Q_{-1}^{-}} \gamma^{\alpha} {1 \over Q_{0}^{-}} \gamma^{\beta} {1 \over
Q_{-2}^{-}}
\gamma^{\gamma} {1 \over Q_{-(23)}^{-}}\right]\,,\nonumber \\
\eea
where $Q_{i}^{\pm}= \rlap/Q_{i} \pm m_{e}$  
($Q_{0}=q,~Q_{23}=q+p_{2}+p_{3}$,~$Q_{2}=q+p_{2},~Q_{-1}=q-p_{1},~{\rm
  etc.}$), \\  
$1/ \Delta_{Z}=1/\left({(p_{4}+p_{5})}^{2}-M_{Z}^{2}\right )\sim-1/M_{Z}^{2}$  
\, and  
$$v_{e}^{\pm}=v_{e}\pm a_{e},\quad v_{e}=-\frac 1 2+2 s_{W}^{2} \,  ,
   \  \ a_{e}=\frac 1 2\ .$$
Furthermore, $s_W$ and $c_W$ are the sine and cosine of the Weinberg
angle,
$w^{\pm}=(1 \pm \gamma_{5})/2$ and 

\bea\Gamma_{\mu}={\bar v}_{+}(5) \gamma_{\mu} w_{-} u_{-}(4)\ .\eea

By reversing the trace in $A_{321}$, changing 
$q$ to $-q$ and $m_e$ to $-m_e$ (only even powers of $m_e$
survive) one finds  that $A_{321}$ is equal to $A_{123}$ with
the replacement $w^{\tau}\rightarrow w^{-\tau}$. Thus, in the sum, 
the $\gamma_5$  contribution cancels in the trace and the pair loses 
memory of the axial part:
\bea \label{as}
A_{123}^{\alpha \beta \gamma}+A_{321}^{\alpha \beta \gamma}=-2{(g
s_{W})}^3 {\left( {g \over 2 c_{W}}
\right)}^2
v_{e}\Gamma_{\mu} {1 \over \Delta_{Z}} L_{1}^{\mu \alpha
\beta \gamma} \nonumber \\
L_{1}^{\mu \alpha \beta \gamma}={1 \over {(2 \pi)}^{4}}  \int dq^{n}
{\rm \,Tr\,}\left[ \gamma^{\mu} {1
\over
Q_{23}^{-}} \gamma^{\gamma} {1 \over Q_{2}^{-}} \gamma^{\beta} {1 \over
Q_{0}^{-}}
\gamma^{\alpha} {1 \over Q_{1}^{-}} \right] \,.
\eea
 
The same trick can be applied to each pair of diagrams
of type $B$. For instance 
\bea
B_{123}^{\alpha \beta \gamma}+B_{321}^{\alpha \beta \gamma}=-4{(g
s_{W})}^3 {\left( {g \over 2 \sqrt{2}}
\right)}^2
 \Gamma_{\mu} 
 L_{2}^{\mu \alpha \beta \gamma} \nonumber \\
\label{l2}
L_{2}^{\mu \alpha \beta \gamma}={1 \over {(2 \pi)}^{4}} \int dq^{n}
{\rm \,Tr\,}\left[ \gamma^{\mu} {1
\over
Q_{23}^{-}} \gamma^{\gamma} {1 \over Q_{2}^{-}} \gamma^{\beta} {1 \over
Q_{0}^{-}}
\gamma^{\alpha} {1 \over Q_{1}^{-}} \right]{1 \over \Delta_{W}(q)}\,,
\eea
where $\Delta_{W}(q)=(q+p_{2}+p_{3}+p_{5})^{2}-M_{W}^{2}$.
At this point, to consistently retain only terms 
${\cal O}(\frac{1}{M_W^2})$,  one has to
expand ${1/ \Delta_{W}(q)}$ as follows:
\bea \label{split}
{1 \over \Delta_{W}(q)}={1 \over {q}^{2}-M_{W}^{2}} -
{k^{2} + 2 q \cdot k \over (q^{2}-M_{W}^{2}) ({(q+k)}^{2}
 -M_{W}^{2})},
\eea
where $k=p_{2}+p_{3}+p_{5}$. Once introduced in eq. (\ref{l2}), 
the first term in the r.h.s. of eq. (\ref{split}) allows the following
splitting 
\bea \label{trick1}
{1 \over q^{2}-m_{e}^{2}} \ \ {1 \over q^{2}-M_{W}^{2}} \sim 
-{1 \over M_{W}^{2}} \left( {1 \over q^{2} -m_{e}^{2}} - {1 \over
q^{2}-M_{W}^{2}} \right) + {\cal O}(1/M_{W}^{4})\,.
\eea
The first term in the previous equation gives an $L_{1}$-type
integral, while we have explicitly checked that, after adding
the contributions of all $B^{ijk}$,
the second term vanishes at order $1/M_{W}^{2}$. 

On the other hand, the second term on the r.h.s. of eq. (\ref{split}) 
is of order $1/M_{W}^{4}$, and can therefore be neglected.
This can be easily seen by splitting again the denominators as shown in 
eq. (\ref{trick1}). That procedure already generates 
an overall $1/M_{W}^{2}$ factor, and what remains is finite and 
proportional to an extra factor $1/M_{W}^{2}$.

In conclusion, at  leading order in $1/M_{W}^{2}$, the  set 
of four diagrams (from a total of 12) is
always proportional to $L_{1}$:
\bea \label{set1}
A_{123}^{\alpha \beta \gamma}+A_{321}^{\alpha \beta \gamma}+
B_{123}^{\alpha \beta \gamma}+B_{321}^{\alpha \beta \gamma}=
-{g^{5} s_{W}^3 \over 2} (1+v_{e}) \Gamma_{\mu}
{1 \over {\Delta_{Z} c_W^{2}}} L_{1}^{ \mu \alpha \beta \gamma}\,.
\eea
Similar results are obtained for the other two groupings of four diagrams,
the only
difference being a trivial change of momenta and indices inside
$L_{1}^{\mu \alpha\beta\gamma}$.

At this point, the correspondence with the four-photon scattering
\cite{book} is evident. In fact, by fixing the fourth photon leg
and calling $C_{123}^{\alpha \beta \gamma}$ the corresponding
amplitude in fig. 1b, one finds a contribution proportional
to the same integral $L_1$:
\bea \label{set2}
C_{123}^{\alpha \beta \gamma}+C_{321}^{\alpha \beta \gamma}
=2 {g^4 s_{W}^{4}}
\epsilon_{\mu}({\overrightarrow{P_4}},\lambda_{4}) 
 L_{1}^{\mu \alpha \beta \gamma}\,, 
\eea
and similar results for the two remaining combinations
$C_{132}^{\alpha \beta \gamma}+C_{231}^{\alpha \beta \gamma}$
and
$C_{213}^{\alpha \beta \gamma}+C_{312}^{\alpha \beta \gamma}$.
Therefore, since the four-photon process is described, at energies
below $m_{e}$, by the Euler--Heisenberg Lagrangian of
eq. (\ref{e-h}), it turns out that the same effective 
Lagrangian can  also be used to describe processes 
(\ref{eq4}), (\ref{eq5}) and (\ref{eq6}).
The only change is the replacement of a photon line
with a neutrino pair. That can be formally achieved by considering the
two neutrinos as a new ``gauge field'' 
\footnote{We are using a different convention for $\gamma_5$, 
with respect to ref. \cite{dic3}.}
${\tilde A}_{\nu}\equiv{\bar \psi} \gamma_{\nu}(1 -
\gamma_{5}) \psi=2 \Gamma_{\nu}$, with field strength ${\tilde F}_{\mu
\nu}$.

In conclusion, the effective Lagrangian for the five-leg interaction
depicted in fig. 3 reads
\bea \label{our}
{\cal L}_{\rm eff}={C \over 180}
\left [ 5\left({\tilde F}_{\mu\nu}F^{\mu\nu}     \right)
          \left(       F _{\lambda\rho} F^{\lambda\rho}\right)
-14 {\tilde F}_{\mu\nu}F^{\nu\lambda}F_{\lambda\rho}F^{\rho\mu}\right ].
\label{proc}
\eea
The constant $C$ remains to be fixed.
This can be easily obtained by considering the following
ratios of amplitudes in the large-$m_e$ limit
\bea
\lim_{\rm {large~m_e}}\ \ {{\cal A}_{4 \gamma}^{\rm SM} \over {\cal
A}_{P}^{\rm SM}}=
{{\cal A}_{4 \gamma}^{\rm eff} \over {\cal A}_{P}^{\rm eff}}\,,
\eea

\vspace*{-1cm}
\bfig(100,120)
\sof(-50,20)
\flin{40,0}{110,0}  \flin{110,0}{180,0}
\glin{110,0}{67,37}{7} 
\glin{110,0}{110,62}{6} 
\glin{110,0}{153,37}{7} 
\GOval(110,0)(4,4)(0){0}     
\Text(65,-5)[t]{$p_4~~~~\nu$}
\Text(155,-5)[t]{$\nu~~~~p_5$}
\Text(115,50)[l]{$\gamma$}
\Text(75,23)[t]{$\gamma$} 
\Text(145,23)[t]{$\gamma$}
\Text(63,45)[r]{$p_1,~\epsilon_1$} 
\Text(155,45)[l]{$p_3,~\epsilon_3$}
\Text(110,70)[b]{$p_2,~\epsilon_2$}
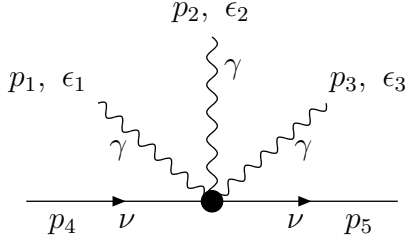
\efig{Fig. 3:  Five-leg photon--neutrino effective interaction.}

where $P$ stands for any of the processes
(\ref{eq4}), (\ref{eq5}) or (\ref{eq6}), and $4\gamma$ is
the four-photon interaction.

The first ratio is given by eqs. (\ref{set1}) and (\ref{set2}): 
\bea \label{rat1}
{{\cal A}_{4 \gamma}^{\rm SM}\over {\cal A}_{P}^{\rm SM}}={4 M_{W}^{2}
s_{W}^{2}
\epsilon_{\mu}({\overrightarrow{P_4}},\lambda_{4}) \over e (1+v_{e})
\Gamma_{\mu} }\,,
\eea
while the ratio between the two effective amplitudes can be
calculated using the effective Lagrangians in eqs. (\ref{e-h}) 
and (\ref{our}):
\bea \label{rat2}
{{\cal A}_{4 \gamma}^{\rm eff}\over {\cal A}_{P}^{\rm eff}}={e^{4}
\epsilon_{\mu}({\overrightarrow{P_4}},\lambda_{4}) \over 8 \pi^{2} C
\Gamma_{\mu}}
\frac{1}{m^4_e}\,,
\eea
where we have used the fact that the amplitudes for 
four-photon and photon--neutrino processes have 
exactly the same momentum dependence in both  SM and effective theory. 
Finally, from eqs. (\ref{rat1}) and (\ref{rat2}) we derive
\bea
C={g^5 s_{W}^{3} (1 + v_{e}) \over 32 \pi^{2} m_{e}^{4} M_{W}^{2}}=\ {2 G_F
\alpha^{3/2}(1 + v_{e}) \over \sqrt{2\pi} m_e^4}\label{prefactor}\,,
\eea
which agrees with the prefactor of the effective Lagrangian used 
in ref.  \cite{dic3}.

From the previous derivation, it is clear that
the analogy between the two processes also holds at the 
level of the exact calculations in the SM, 
within the approximations discussed. Therefore, by substituting 
\bea
C \rightarrow {\alpha^2 \over m_{e}^{4}}  \qquad {\rm and}
\qquad
\Gamma_{\mu} \rightarrow 2
\epsilon_{\mu}({\overrightarrow{P_4}},\lambda_{4})  
\eea
(for any polarization $\lambda_{4}=\pm$)
in the five-leg photon--neutrino process,
the amplitudes of the four-photon process \cite{book} should 
always be recovered. Notice that in the equivalence between our process $P$ and
the four-photon process, one external leg of the four-photon amplitude should
be taken off-shell. 
 
The previous derivation can be generalized in a straightforward way to
an arbitrary  number of external photons.

\section{Polarized and total cross sections 
for $\gamma\gamma\to \gamma \nu\overline\nu$ }\label{cross}

In this section, we give details for the computation of 
process (\ref{eq5}). Our main motivation is the  
disagreement between the calculations of refs. \cite{dic3} and \cite{shab}.
For massless neutrinos, the differential cross section reads
\bea\begin{array}{ll}
d\sigma = {1\over 4 (P_1\cdot P_2)}{ |T_{fi}|^2 } 
d \tilde{P}_3 d \tilde{P}_\nu d \tilde{P}_{\bar{\nu}} (2\pi)^4
\delta^4(P_i-P_f) \  \\ 
P_i= P_1 +P_2\,,~P_f= P_3+P_\nu+P_{\bar{\nu}}\,,~ 
 d \tilde{P}={d^3 P\over (2\pi)^3 2\omega}\,.\end{array}
\eea 
Here, $P_{1,2}$ are the momenta of the incoming photons,
$P_3$ the momentum of the outcoming one and 
$P_{\nu,\bar\nu }$ the 
momenta of the outcoming neutrino-antineutrino pair.
Therefore the dilepton energy is 
\bea
k_4 \equiv P_\nu+P_{\bar{\nu}}\label{k4}\,;
\eea
$T_{fi}$ reads
\bea
T_{fi}= C M_\sigma \bar{u}_\nu(P_\nu)\gamma^\sigma
(1-\gamma_5)u_{\bar{\nu}}(P_{\bar{\nu}})\,,
\eea
then 
\bea
|T_{fi}|^2 = 4 C^2 
\left[(P_\nu\cdot M) (P_{\bar{\nu}}\cdot M^*) +(\nu\leftrightarrow \bar{\nu}) -
(P_{\bar{\nu}}\cdot P_\nu)|M|^2 -  
i \varepsilon_{\mu\sigma\rho\eta} 
P_\nu^\mu P_{\bar{\nu}}^\sigma M^\rho M^{* \eta}\right]\label{tfi2}\,,
\eea
with $C$ given in eq. (\ref{prefactor}) and 
\bea
M_{\sigma}\equiv \
 \epsilon^{\alpha_1}(\vec P_1,\lambda_1)\epsilon^{\alpha_2}
(\vec P_2,\lambda_2) 
\epsilon^{\alpha_3}(\vec
P_3,\lambda_3)M_{\alpha_1\alpha_2\alpha_3\sigma}(P_1, P_2, P_3, k_4)\,.
\label{I} 
\eea
The functions $\epsilon(\vec P_i,\lambda_i)$ are the polarization vectors
of the  
photons and $M_{\alpha_1\alpha_2\alpha_3\sigma}$ is the scattering 
tensor defined by
\bea
M^{\alpha_1\alpha_2\alpha_3\sigma}(P_1, P_2, P_3, k_4)\equiv \ \ 
U^{\alpha_1 \beta_1 \lambda_1 \alpha_2 \beta_2 \lambda_1 \alpha_3 \beta_3 
\lambda_3 }\ 
T_{\lambda_1\beta_1 \lambda_2\beta_2\lambda_3\beta_3\mu\nu} \ 
S^{\mu\nu\rho\sigma}\  k_4^\rho\,.
\label{I4} 
\eea
Finally, the tensors $U$, $T$, and $S$ are 
\bea \begin{array}{ll}
U^{\alpha_1 \beta_1 \lambda_1 \alpha_2 \beta_2 \lambda_2 \alpha_3 \beta_3 
\lambda_3 }
\equiv |{\epsilon_{ijk} }|P_i^{\lambda_1} P_j^{\lambda_2}
 P_k^{\lambda_3} 
g^{\alpha_i\beta_1}g^{\alpha_j\beta_2}g^{\alpha_k\beta_3}\\ 
S_{\mu\nu\rho\sigma}  \equiv \left(g_{\mu\rho}g_{\nu\sigma}
\ - (\mu \leftrightarrow \nu)\right)\\ 
T^{\lambda_1\beta_1 \lambda_2\beta_2\lambda_3\beta_3\mu\nu} \equiv
-{10\over 180} \left\{\left [ g^{\mu\lambda_1}g^{\nu\beta_1}\left
( g^{\lambda_2\lambda_3}g^{\beta_2\beta_3}-(\lambda_3\leftrightarrow \beta_3)
\right ) \right]
- \left [\mu\leftrightarrow \nu\right]\right\} \\ 
 +{14\over 180} \left\{\left[\left (g^{\lambda_1\nu}g^{\beta_1\lambda_2}-
 (\beta_1\leftrightarrow \lambda_1) \right)
 \left (g^{\beta_2\lambda_3}g^{\beta_3\mu}-(\beta_3\leftrightarrow \lambda_3) 
 \right)\right]-
\left[\beta_2\leftrightarrow \lambda_2\right]\right\}\,.\end{array}
\label{tensors} 
\eea
Using the Ward identity  $k_4^\mu M_\mu=0$  and the fact that the last 
term of eq. (\ref{tfi2}) cancels\footnote{Since the tensor $M^{\alpha_1\alpha_2\alpha_3\sigma}$ 
in eq. (\ref{I4}) is real by construction, the last
term of eq. (\ref{tfi2}) immediately vanishes, for symmetry reasons, when 
taking a real representation for the polarization vectors
appearing in eq. (\ref{I}).}
when taking the square, we simply get 
\bea
|T_{fi}|^2= -4 C^2 \left[2( P_\nu\cdot  M) (P_\nu\cdot M^*)  +
(P_{\bar{\nu}}\cdot P_\nu)|M|^2 \right]\label{contrib}\,.
\eea

To compute the phase-space integrals, 
we work in the centre-of-mass frame of the two incoming photons:
\bea
P_1=(\omega,\vec \omega)=\omega(1,\vec n), \quad P_2=\omega(1,-\vec n),
\quad P_3=(\omega_1,\vec \omega_1)=\omega_1(1,\vec{n}'),
\nn \\ P_\nu=(\omega_2,\vec \omega_2)=\omega_2(1,\vec{n}'') ,
\quad P_{\bar\nu}=(2\omega-\omega_1 -\omega_2 ,-\vec\omega_1   -\vec
\omega_2 ),
\eea 
where  $\vec n,\ \vec{n}'$ 
and $\vec{n}''$ are unit vectors.
In this process, $P_\nu$  and $P_{\bar{\nu}}$ 
always appear as a sum in the amplitude, and
$k_4=(2\omega -\omega_1, -\vec\omega_1 )$.

The scattering plane is defined by the unit vectors 
$\vec n$ and $\vec{n}'$. The polarization vectors of the two photons 
with momentum $\vec \omega$ and $\vec \omega_1$ can be defined with
respect to this plane, as follows:
\bea
\epsilon(\vec n, \perp) &=& \epsilon(\vec{n}', \perp)=
\left(0,\varepsilon_{ijk}{n^jn'^k\over \sin\theta}\right)
\nn\\
\epsilon(\vec n, \parallel)&=&
\left(0,{\vec{n}'  - \cos \theta \ \vec n\over \sin\theta}\right)
\nn\\
\label{polari}
\epsilon(\vec{n}',\parallel)&=&
\left(0,{\vec{n}'\cos \theta-\vec n \over \sin\theta}\right)\,,~~~~~~~~
 \cos\theta= \vec n\cdot\vec{n}'\,. 
\eea
The completeness and  normalization relations are easily checked
\bea
&&\sum_{\lambda=\perp,\ \parallel}\epsilon^\mu(\vec k,\lambda)
\epsilon^\nu(\vec k,\lambda)= -g^{\mu \nu} + {k^\mu
k'^\nu+k'^\mu
k^\nu\over (k\cdot k')}\nn  \\
&&\epsilon^\mu(\vec k,\lambda)\epsilon_\mu(\vec k, \lambda)=-1\ , \quad {\mathrm{for }}\, \, \lambda
=\perp,\parallel\,.  
\label{completeness}
\eea
In the previous equations, $k=(k^0,\vec k)$ is the momentum
of the photon and $k'$ any arbitrary
vector non-parallel to $k$ and such that ${k'}^2=0$. The simplest choice
is
$ k'\equiv(-k^0,\vec k)$.
In the frame defined above, by using the fact that $M$ 
is real with our choice for the polarization vectors, 
each of the two terms in eq. (\ref{contrib}) gives  a contribution
proportional to $|M|^2$, for each set of polarizations 
$(\lambda_1,\lambda_2,\lambda_3)$.
After integrating over the lepton pair, the polarized cross section 
has the form
\bea
d\sigma(\lambda_1,\lambda_2,\lambda_3) = 
{G_F^2 a^2 \alpha^3\over 3\ m_e^8 (P_1\cdot P_2)( 2\pi)^5}
M^2(\lambda_1,\lambda_2,\lambda_3) k_4^2 {d^3{P}_3\over
\omega_1}
\label{difcro}
\,, 
\eea
where $a= 1+v_e$.
By inserting eq. (\ref{polari}) in eq. (\ref{difcro}), one explicitly gets
\bea 
{d\sigma (\perp,\perp,\perp) \over \sin\theta d \theta d\omega_1}&=&
8{G_F^2 a^2 \alpha^3\over 6075\ m_e^8 \pi^4 }
\,{\omega^3}\,{{{\omega_1^3}}} (\omega-\omega_1)\,
  {{\left( 4\,\omega  -\sin^2\theta \,
       {\omega_1} \right) }^2}
 \nn \\ 
{d\sigma ((\parallel,\parallel,\parallel)) 
\over \sin\theta d \theta d\omega_1}&=& 
8{G_F^2 a^2 \alpha^3\over 6075\ m_e^8 \pi^4 }
{ \omega^3}\,{{ { \omega_1^3}}}(\omega-\omega_1)\,
  \left( -4\,\left( -4 + 3\,{{ {\sin^2 \theta}}} \right) \,{ \omega^2}
  \right.\nn\\ &+&
    \left. 4\,{{ {\sin^2\theta}}}\, \omega\, { \omega_1} + 
    {{ {\sin^4\theta}}}\,{{ { \omega_1}}^2} \right) 
\nn\\ 
{d\sigma  (\perp,\parallel,\perp)\over \sin\theta d\theta d\omega_1}&=&
 {G_F^2 a^2 \alpha^3\over 48600\ m_e^8 \pi^4 }
 {\omega^3}\,{{ {\omega_1^3}}}(\omega-\omega_1)\,
  \left( 4\,\left( 410 + 374\, {\cos\theta} - 93\,{{ {\sin^2\theta}}}
       \right) \,{\omega^2}\right.\nn\\
       &-& \left.4\,\left( 79 + 33\, {\cos\theta} \right) \,
     {{ {\sin^2\theta}}}\,\omega\, {\omega_1} + 
    9\,{{ {\sin^4\theta}}}\,{{ {\omega_1}}^2} \right) 
\nn \\ 
 {d\sigma  (\perp,\perp,\parallel)\over \sin\theta d\theta d\omega_1}&=&
 {G_F^2 a^2 \alpha^3\over 12150\ m_e^8 \pi^4 }
 \,{ \omega^3}\,{{ { \omega_1^3}}}(\omega-\omega_1)\,
  \left( 4\,\left( 9 + 7\,{{ {\sin^2\theta}}} \right) \,{ \omega^2}  \right.
  \nn\\ &-&
    \left.  
    112\,{{ {\sin^2\theta}}}\, \omega\, { \omega_1} + 
    49\,{{ {\sin^4\theta}}}\,{{ { \omega_1}}^2} \right) 
\nn\\
 {d\sigma  (\parallel,\parallel, \perp)\over \sin \theta d\theta d\omega_1}&=&
 {G_F^2 a^2 \alpha^3\over 12150\ m_e^8 \pi^4 }\,{ \omega^3}\,{{ { \omega_1^3}}}
 (\omega-\omega_1)\,
  {{\left( 6\, \omega - 7\,{{ \sin\theta}^2}\, { \omega_1} \right) }^2}
\nn\\
 {d\sigma (\parallel,\perp, \parallel)\over \sin\theta d\theta  d\omega_1}&=&
 {G_F^2 a^2 \alpha^3\over  48600\ m_e^8 \pi^4 }{\omega^3}\,{{
 {\omega_1}}}(\omega-\omega_1)
  {{\left( \left( 34 + 22\, \cos \theta \right) \,\omega  - 
       3\,{{ \sin\theta}^2}\, {\omega_1 } \right) }^2}
\nn\\ 
{d\sigma (\perp, \parallel, \parallel)\over \sin\theta d\theta d\omega_1}&=&
{G_F^2 a^2 \alpha^3\over 48600\ m_e^8 \pi^4 }{\omega^3}\,
{{ {\omega_1^3}}}(\omega-\omega_1)
  {{\left( \left( 34 - 22\, \cos \theta \right) \, \omega - 
       3\,{{ \sin\theta}^2}\, { \omega_1} \right) }^2}
\nn\\ 
{d\sigma (\parallel,\perp, \perp)\over \sin \theta d\theta d\omega_1}
&=&{G_F^2 a^2 \alpha^3\over 48600\ m_e^8 \pi^4 }{\omega^3}\,{{ {\omega_1^3}}}
(\omega-\omega_1)
  \left( 4\,\left( 317 - 374\, {\cos\theta} + 93\,{{ {\cos^2\theta}}}
       \right) \,{ \omega^2} \right.\nn\\
       &-& \left.4\,\left( 79 - 33\, {\cos\theta} - 
       79\,{{ {\cos^2\theta}}} + 33\,{{ {\cos^3\theta}}} \right) \, \omega\,
      { \omega_1}  + 9\,{\sin^4\theta}\,
     {{ { \omega_1}}^2} \right)\,. 
\eea
By summing the above contributions and averaging over 
the initial photon polarizations, one gets the unpolarized 
differential cross section
\bea 
 {d\sigma \over \sin \theta d\theta d\omega_1}&=&  
{G_F^2 a^2 \alpha^3\over 48600\ m_e^8 \pi^4 }{\omega^3}{\omega_1^3}
(\omega-\omega_1)
  \left( 2224{\omega^2} -592{{ \sin^2\theta}}\,{\omega^2} \right.\nn \\ &-& 
   \left.  520\,{{ \sin^2\theta}}\,\omega\, {\omega_1} +
    139\,{{ \sin^4\theta}}\,{{ {\omega_1}}^2} \right)\,, 
\label{dsigma}\eea
from which the total cross section immediately follows
\bea
\sigma(\gamma\gamma\to \gamma\nu\bar\nu)={2144  \over 637875}
{G_F^2 a^2 \alpha^3\over  \pi^4 }
\left({\omega\over m_e}\right)^8 \omega^2\,.
\label{resint}
\eea
The formulae in eqs. (\ref{dsigma}) and (\ref{resint}) agree with the 
results presented in ref. \cite{dic3} and
disagree with the ones obtained by Hieu and Shabalin \cite{shab}.

As a cross-check, we computed the polarized amplitudes for the process
$\gamma \gamma \to \gamma \gamma $, using the effective
Lagrangian of eq. (\ref{our}), where we substituted back 
the neutrino current with a photon polarization, and put $\omega=\omega_1$ in
order to
have an on-shell photon.
We recovered all the polarized amplitudes given in ref. \cite{book}
\footnote{In  \cite{book}, the polarized amplitudes 
are correctly reported, while we found a mistake in the general
expression for the 
four-photon amplitude. 
The global sign of the fifth line of eq. (54.21) should be a minus.
Since the computation of ref. \cite{shab} is 
based on ref. \cite{book}, that mistake could 
be the source of the error in ref. \cite{shab}.}.

\section{Cross section for 
$\gamma\nu\to\gamma\gamma\nu$}\label{directe}
In this section, we present results for the differential and the
total cross sections of process (\ref{eq4}). 
As a cross-check, two independent calculations have been performed. 
On the one hand, we directly squared the amplitude with the help of 
the completeness relation of eq. (\ref{completeness}). On the other
hand, we separately calculated all helicity amplitudes. 
The corresponding polarized cross sections are given in the appendix.  
The unpolarized differential cross section reads, in both cases
\bea
{d^2\sigma\over d\omega_1d\omega_2}
&=&{a^2\alpha^3 G_F^2\omega^3\over 60750\pi ^4 m_e^8}
     \left( 7588\omega^5 - 21055\,\omega^4
        \left(  \omega_1 +  \omega_2 \right) \right. \nn\\ &+&
  \left.2\omega^3\,\left( 11063\,\omega_1^2 + 
          21750\, \omega_1\, \omega_2 + 11063\,\omega_2^2
           \right) \right.\nn \\&-& \left.2\,\omega^2\,
        \left( 6067\,\omega_1^3 + 
          15552\,\omega_1^2\, \omega_2 + 
          15552\, \omega_1\,\omega_2^2 +
          6067\,\omega_2^3 \right) \right. \nn\\ &+&  
    \left.   2\,\omega\,\left( 2085\,\omega_1^4 + 
          5372\,\omega_1^3\, \omega_2 + 
          8103\,\omega_1^2\,\omega_2^2 + 
          5372\, \omega_1\,\omega_2^3 + 
          2085\,\omega_2^4 \right)  \right.\nn\\ &-& 
   \left.    139\,\left( 5\,\omega_1^5+ 
          15\,\omega_1^4\, \omega_2 + 
          26\,\omega_1^3\,\omega_2^2 + 
          26\,\omega_1^2\,\omega_2^3 + 
          15\, \omega_1\,\omega_2^4 + 5\,\omega_2^5
           \right)  \right) \label{partial}\,,
\eea
where $\omega_1$ and $\omega_2$ are the final photon energies,
computed in the centre-of-mass of the initial state. 
Integrating over $\omega_1$ and $\omega_2$ gives the total cross section
\bea
\sigma(\gamma\nu\to\gamma\gamma\nu)=\int_0^\omega d\omega_1
\int_{(\omega-\omega_1)}^\omega d\omega_2\,  
\frac{d^2\sigma}{d\omega_1 d\omega_2}=
{262 \over 127575}{G_F^2 a^2 \alpha^3\over  \pi^4 }
\left({\omega\over m_e}\right)^8 \omega^2\label{tot}\,.
\eea
The results in eqs. (\ref{partial}) and (\ref{tot}) agree 
with the corresponding expressions in ref. \cite{dic3}, which
we confirm.

\section{Conclusions}\label{remarks}
In this work, we justified the effective Lagrangian approach, 
 based on the four-photon Euler--Heisenberg Lagrangian \cite{EH}, 
to compute five-leg photon--neutrino processes.
We gave all essential steps for the derivation of the effective
Lagrangian at  leading order in the Fermi theory. 
We computed the processes
$\gamma\gamma\to\gamma\nu \bar{\nu}$ and $\gamma\nu\to\gamma\gamma\nu$,  
explicitly listing all polarized cross sections. 

Concerning the process $\gamma\gamma\to\gamma\nu \bar{\nu}$, we confirm the
results of the calculation reported in ref. \cite{dic3}, while we disagree 
with the expressions of ref. \cite{shab}.

\section*{Acknowledgements}
We thank S. Bertolini, G. F. Giudice, A. Masiero and  P. Nason 
for helpful remarks. J.M. acknowledges financial support from a
Marie Curie EC Grant (TMR-ERBFMBICT 972147).

\clearpage
\appendix 
\begin{center}{\bf Appendix A 

\bf  Polarized cross sections for the 
$\gamma\nu\to \gamma\gamma\nu$ process}\label{polar}
 \end{center}
\noindent 
We work in the centre-of-mass frame of the photon--neutrino initial
state:
\bea
&& P_1=(\omega,\vec \omega)=\omega(1,\vec n), \quad P_2=
(\omega_1,\vec \omega_1)=\omega_1(1,\vec{n}'),
\quad P_3=(\omega_2,\vec \omega_2)=\omega_2(1,\vec{n}''),\nn 
\\ && P_4=(\omega,-\vec \omega)=\omega(1,-\vec n) ,
\quad P_5=(2\omega-\omega_1 -\omega_2 ,-\vec\omega_1   -\vec \omega_2 )\,,
\eea 
where  $\vec n,\vec{n}'$ 
and $\vec{n}''$ are unit vectors, $P_1$, $P_2$ and $P_3$ are the initial
and the two final photon momenta, respectively, while  $P_{4/5}$  
are the initial/final neutrino momenta. 
The five independent phase-space variables can be chosen as 
 the two energies
 $\omega_{1,2}$ and three angles that determine the overall
 orientation of the tripod 
  $(\overrightarrow{P_2},\overrightarrow{P_3},\overrightarrow{P_5})$. Two
 angles $(\alpha_1,\phi_1)$ fix the direction of $\overrightarrow{P_2}$ 
  and one angle $\phi$ takes care of the rotation of the  
 $(\overrightarrow{P_3},\overrightarrow{P_5})$ system around
 $\overrightarrow{P_2}$.
 With this choice of variables, the polarization vectors for the incoming
 photon read
 \bea \epsilon(\vec n, \parallel)=
 \left (\begin{array}{l}
 0\\
 -\sin\phi_1\\-\cos\phi_1\cos\alpha_1\\
 \cos\phi_1\sin\alpha_1 \end{array}\right), \qquad \epsilon(\vec n, \perp)=
 \left (\begin{array}{l}
 0\\
 \cos\phi_1\\-\sin\phi_1\cos\alpha_1\\
 \sin\phi_1\sin\alpha_1 \end{array}\right)\,.
\eea
Denoting by $\theta$ the angle between the two final-state photons 
($\cos\theta=\vec{n}'\cdot \vec{n}''$), our choice for the polarization
vectors of the two outgoing photons is instead
\bea
&&\epsilon(\vec{n}', \perp)=\epsilon(\vec{n}'', \perp)=
\left(0,\varepsilon_{ijk}{n'^j n''^k\over \sin\theta}\right)
\nn\\
&&\epsilon(\vec{n}', \parallel)=
\left(0,{\vec{n}''  - \cos \theta\vec{n}'\over \sin\theta}\right)\ \ , 
\quad \epsilon(\vec{n}'',\parallel)=
\left(0,{ \vec{n}''\cos \theta-\vec{n}' \over \sin\theta}\right)\,.
\eea
The computation of each set of polarizations gives
the following polarized cross sections:
\bea
{d^2\sigma(\perp,\perp,\perp)\over d\omega_1d\omega_2}&=&
{d^2\sigma(\parallel,\perp,\perp)\over d\omega_1d\omega_2}=
{a^2\alpha^3 G_F^2\omega^3\over 243000\pi ^4 m_e^8}
\left( 6598\omega ^5 - 365 \omega_1^5 - 
       853 \omega_1^4  \omega_2 \right. \nn \\ &-&
       2822 \omega_1^3 \omega_2^2 - 
       2822 \omega_1^2 \omega_2^3 - 
       853 \omega_1 \omega_2^4 - 
       365 \omega_2^5  \nn \\&-& 
       14477\omega ^4\left(  \omega_1 +  \omega_2 \right)  + 
       \omega ^3\left( 10378 \omega_1^2 + 
          28144 \omega_1 \omega_2 + 10378 \omega_2^2
           \right)  \\  &-&  2\omega ^2
        \left( 2041 \omega_1^3 + 
          8677 \omega_1^2 \omega_2 + 
          8677 \omega_1 \omega_2^2 + 
          2041 \omega_2^3 \right)  \nn \\ &+& \left.  
       4\omega \left( 487 \omega_1^4 + 
          1135 \omega_1^3 \omega_2 + 
          2858 \omega_1^2 \omega_2^2 + 
          1135 \omega_1 \omega_2^3 + 
          487 \omega_2^4 \right)  \right) \nn
\eea

\bea
{d^2\sigma(\parallel,\parallel,\parallel)\over d\omega_1d\omega_2}&=&
{d^2\sigma(\perp,\parallel,\parallel)\over d\omega_1d\omega_2}=
{a^2\alpha^3 G_F^2\omega^3\over 243000\pi ^4 m_e^8}
  \left( 3342\omega^5 - 365\omega_1^5 - 
       677    \omega_1  ^4   \omega_2 \right. \nn \\ &-& 
       1238    \omega_1  ^3     \omega_2  ^2  - 
       1238    \omega_1  ^2     \omega_2  ^3  - 
       677  \omega_1     \omega_2  ^4  - 
       365    \omega_2  ^5  \nn \\&-& 
       9373 \omega^4 \left(   \omega_1  +   \omega_2  \right)  + 
        \omega^3 \left( 9762    \omega_1  ^2  + 
          22600  \omega_1   \omega_2  + 9762    \omega_2  ^2 
           \right) \\  &-& 14 \omega^2 
        \left( 367    \omega_1  ^3  + 
          1183    \omega_1  ^2   \omega_2  + 
          1183  \omega_1     \omega_2  ^2  + 
          367    \omega_2  ^3  \right)  \nn \\ &+& \left. 
       4\,\omega\left( 443    \omega_1  ^4  + 
          1003    \omega_1  ^3   \omega_2  + 
          2418    \omega_1  ^2     \omega_2  ^2  + 
          1003  \omega_1     \omega_2  ^3  + 
          443    \omega_2  ^4  \right)  \right) \nn
     \eea

\bea
{d^2\sigma(\perp,\parallel,\perp)\over d\omega_1d\omega_2}&=&
{d^2\sigma(\parallel,\parallel,\perp)\over d\omega_1d\omega_2}=
{a^2\alpha^3 G_F^2\omega^3\over 243000\pi ^4 m_e^8}  
 \left( 
 10206\omega^5 - 1025\omega_1^5 - 3273\omega_1^4 \omega_2\right. 
 \nn \\&-& 
       5066\omega_1^3\omega_2^2 - 
       5330\omega_1^2 \omega_2^3 - 
       3537\omega_1\omega_2^4 - 
       1025\omega_2^5\nn 
        \\  &-& 
       \omega^4\left(
        30449\omega_1 + 29921\omega_2
         \right)  + 
       2\omega^3
       \left(
        17289\omega_1^2 + 
          30814\omega_1\omega_2 + 16893\omega_2^2
           \right) 
           \\  
           &-&
            2\omega^2
        \left( 
        9829\omega_1^3 + 22757\omega_1^2 \omega_2 + 
          22493\omega_1\omega_2^2 + 
          9829\omega_2^3 
          \right) \nn\\  &+& \left.
       4\omega\left( 
       1587 \omega_1^4 + 
          4270\omega_1^3 \omega_2 + 
          5465\omega_1^2\omega_2^2 + 
          4336\omega_1\omega_2^3 + 
          1653\omega_2^4
           \right)  
          \right) \nn
\eea

\bea
{d^2\sigma(\perp,\perp,\parallel)\over d\omega_1d\omega_2}&=&
{d^2\sigma(\parallel,\perp,\parallel)\over d\omega_1d\omega_2}=
{a^2\alpha^3 G_F^2\omega^3\over 243000\pi ^4 m_e^8}
\left(10206\omega^5 - 1025\omega_1^5 - 
       3537\omega_1^4 \omega_2  \right. \nn\\
       &-&5330\omega_1^3\omega_2^2 - 
       5066\omega_1^2\omega_2^3 - 
       3273 \omega_1\omega_2^4 - 
       1025\omega_2^5
 \nn \\
       &-&\omega^4\left( 29921\, \omega_1 + 30449\, \omega_2 \right)  + 
       2\omega^3\left( 16893\omega_1^2 + 
          30814\omega_1\, \omega_2 + 17289\omega_2^2
           \right) \\ & -& 2\omega^2
        \left( 9829\omega_1^3 + 
          22493\omega_1^2\, \omega_2 + 
          22757\, \omega_1\omega_2^2 + 
          9829\omega_2^3 \right)  \nn \\ &+&
             \left.   4\omega\left( 1653\omega_1^4 + 
          4336\omega_1^3\, \omega_2 + 
          5465\omega_1^2\omega_2^2 + 
          4270 \omega_1\omega_2^3 + 
          1587\omega_2^4 \right)  \right)\,. \nn
\eea

\end{document}